\documentclass[aps,showpacs,twocolumn,toolkits]{revtex4}
\usepackage{graphicx}
\newcommand{\Dslash}{D \! \! \! \! \! \!  \ /}

\begin{document}

\title{Effective chiral restoration in the hadronic spectrum and QCD }

\author{Thomas D. Cohen}

\email{cohen@physics.umd.edu}

\affiliation{Department of Physics, University of Maryland,
College Park, MD 20742-4111}

\begin{abstract}
Effective chiral restoration in the hadronic spectrum has been
conjectured as an explanation of multiplets of nearly degenerate
seen in highly excited hadrons.  The conjecture depends on the
states being insensitive to the dynamics of spontaneous chiral
symmetry breaking. A key question is whether this concept is well
defined in QCD.  This paper shows that it
 is by means of an explicit  formal construction.
This construction allows one to characterize this sensitivity for
any observable calculable in QCD in Euclidean space via a
functional integral. The construction depends on a generalization
of the Banks-Casher theorem. It exploits the fact that {\it all}
dynamics sensitive to spontaneous chiral symmetry breaking
observables in correlation functions arise from fermion modes of
zero virtuality (in the infinite volume limit), while such modes
make {\it no} contribution to any of the dynamics which preserves
chiral symmetry.  In principle this construction can be
implemented in lattice QCD.  The prospect of a practical lattice
implementation yielding a direct numerical test of the concept of
effective chiral restoration is discussed.
\end{abstract}

\maketitle

\section{Introduction}

A striking feature of the hadron spectrum has a number of cases in
which states of the same spin and opposite parity are nearly
degenerate.  One suggestion is that this phenomenon is the result
of ``effective restoration'' of chiral symmetry in the
spectrum\cite{G1,CG,G2,G3,G4}.  The notion that  chiral
multiplets may be relevant to hadron spectrum  in the context of
simple phenomenological models was discussed in ref.~\cite{Jido}.
If this conjecture is correct the spectrum should consist of
approximate chiral multiplets rather than simply have parity
doublets\cite{CG}.  It has been argued that there are indications
in the hadron spectrum for such patterns. The basic theoretic
idea underlying this conjecture is that if the properties of the
high-lying hadrons are largely insensitive to the dynamics of
spontaneous chiral symmetry breaking, then to good approximation
the states will fall into chiral multiplets. The logic is simple:
although the system is in the Nambu-Goldstone phase,
 the particular
states are insensitive to this fact and the spectra act to good
approximation as if they were in the Wigner-Weyl phase.  There is
no theoretical reason why this is excluded, and a simple
calculable model illustrates how the phenomenon could come
about\cite{CG2}. Moreover, there are some general arguments as to
why this should not be regarded as
implausible\cite{CG,G2,G3,G4,Sh}.  At the same time, it should be
noted that such an interpretation remains
controversial\cite{JPS,JPS2} and the empirical evidence from the
spectrum from the spectrum is probably better characterized as
being suggestive rather than compelling.  While there is
additional support for the phenomenon based on the small size of
the couplings of excited hadrons to pions\cite{CN}, the overall
phenomenological evidence is not conclusive.

Given this somewhat unsatisfactory situation, it is important to
ask whether there is any theoretical method to determine if
effective chiral restoration occurs to good approximation in the
spectrum of QCD. There are actually two related issues here. The
first is one of principle: Can one formulate a calculation in QCD
which, if done,
 determines whether effective chiral restoration
occurs to good approximation? As we will demonstrate, the answer
to this question is ``yes''.  The second issue is more practical:
namely, is there a viable method to actually implement such a
calculation? As will be discussed below, the answer to this
question depends at minimum on the existence of reliable lattice
calculations for the masses of excited hadronic resonances. The
problem of determining whether excited resonant states exhibit
effective chiral restoration from the lattice may require
substantially more computational resources than
 simply determining the masses as it requires enough numerical
power to approach multiple limits in a particular order.
Nevertheless, there is a real prospect of using lattice data (at
some future date) to determine whether or not effective chiral
restoration is, in fact,
 responsible for the degeneracies seen in
the spectrum for states of opposite parity.

The crux of the theoretical issue is that the effective chiral
restoration depends on the masses of the excited hadrons be largely
insensitive to the physics of spontaneous chiral symmetry breaking.
To proceed further it is essential to formulate more precisely what
this means.  In the context of simple mean field models such as the
one considered in ref.~\cite{CG2}, it is easy to answer this---there
is a single chiral order parameter in the models and its size is
controlled directly from an external parameter. Hence, in the model
one can directly test how the mass varies with the order parameter.
One would like to do the same thing with QCD. However, there are
problems with this.  In the first place we can not independently
alter the chiral condensate.  Of course, one can add something to
the theory (say, a chemical potential) which alters the chiral
condensate and ask how much the hadron's mass changes.  The
difficulty with this is that we do not know how much of the change
in the hadron's mass is due to the changing condensate and how much
is due to other effects associated with adding something new to the
theory. Moreover, the chiral condensate is not the only chiral order
parameter.

Thus, to proceed we need some algorithm by which we can turn off the
effects of spontaneous chiral symmetry breaking on observables in
QCD without  altering the dynamics not associated with chiral
symmetry breaking.  Given such an algorithm one can calculate the
hadron mass with the effects of spontaneous chiral symmetry breaking
included and excluded, and directly compare the differences.

The first thing to note is that all physical processes involving
hadrons are ultimately describable in terms of correlations
functions of gauge invariant currents.  The correlation functions
may be regarded as the response of the system given certain types
of probes which are determined by the currents. Thus, if one can
find two sets of currents, one of which probes the system in the
conventional way and a second set which are identical in all
respects except that they do not probe the chiral symmetry
breaking aspects of the system, in effect one would be able to
turn off the dynamics of chiral symmetry breaking.  In doing so
it is important to note that the correlation functions are
describable in terms of dispersion relations with the information
contained in spectral functions. The key issue is whether the
spectral functions in the regions of interest are dominated by
the dynamics which preserve chiral symmetry.

The goal is to isolate the dynamics of chiral symmetry breaking
inside the calculation. While it is not clear how to formulate such
an algorithm in Minkowski space, there is a rather straightforward
way to do so in Euclidean space. The key hint in constructing such
currents is an old observation of Banks and Casher\cite{BC} that the
chiral condensate in a Eulclidean formulation is directly
proportional to the density of states of the Dirac operator at zero
virtuality:
\begin{equation}
\langle \overline{q} q \rangle = - \pi \rho (0) \; .\label{bc}
\end{equation}

The Banks-Casher relation applies only to the chiral condensate.
However, the result applies far more generally. In the appropriate
chiral and infinite volume limits (with $V \rightarrow \infty$
taken first) {\it all} effects of chiral symmetry breaking on
{\it any} Eulcidean space correlation function are directly
attributable to the contribution of fermion modes associated with
the external currents that have zero virtuality.  Moreover, these
modes that contribute {\it only} to dynamics which break chiral
symmetry. By comparing correlations functions calculated with
zero modes included in the external currents with analogous
calculations excluding them, one has a direct measure of the role
of chiral symmetry breaking on the Euclidean correlation function.

As will be discussed in detail below, this is a highly nontrivial
numerical problem in practice. However, some insight into this
behavior has already been extracted from lattice simulations by
Degrand\cite{Dg}, who showed that the near zero modes contribute
strongly to the correlation functions at long times but had a
very small contribution to the short-distance behavior.  While
calculations of this sort are not suitable for demonstrating
effective chiral restoration in the spectrum they provide a
useful consistency check.  The results of ref. \cite{Dg} are
consistent with the notion of effective chiral restoration.

For simplicity this paper will only consider non-strange systems,
so the relevant chiral symmetry is $SU(2)_L \times SU(2)_R$ and
will focus on the chiral limit of zero quark mass. The paper is
organized as follows: In the next section the issue of how to
obtain the masses of resonant states from correlation functions
is addressed briefly. The subsequent section demonstrates that in
Euclidean correlation functions the zero virtuality modes
contribute to effects which spontaneously break chiral symmetry
and only to such effect. Next an explicit construction is made
for correlation functions removing the effects of dynamical
symmetry breaking. Following this is a section discussing the
role of the $U(1)$ axial anomaly in the preceding analysis.
Finally, there is a section which addresses the prospects for
implementing it on the lattice.

\section{Hadronic resonances from Euclidean space correlation
functions}

Dispersion relations give both Euclidean space and Minkowski
space correlation functions from the same spectral function. Thus,
if one can deduce the spectral function from a Euclidean space
calculation, then one knows the Minkowski space amplitudes.  Since
all observable properties of hadronic states and their
interactions can be extracted from Minkowski correlation
functions, all one needs is an algorithm to extract the
appropriate spectral function.

\begin{figure}[htb]
\includegraphics[scale=.5]{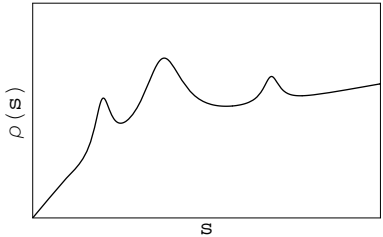}
\caption{\label{fig:res} Caricature of a spectral function
containing three clear resonant structures}
\end{figure}

As a simple example, note that prominent resonances may appear
directly in the spectral function for a two-point correlation
function. Ignoring inessential complications of spin and flavor, the
Euclidean space two-point correlator for a local gauge invariant
current $J$ is given by:
\begin{equation}
\Pi (x)  \equiv \langle J^\dagger (x) J(0) \rangle = \int d s \,
\rho(s) G(x;s)
\end{equation}
where $G(x ;s)$ is the scalar propagator for a point particle of
mass $\sqrt{s} $:
$$G(x;s) = \int \frac{d^4 Q}{(2 \pi)^4}\, \frac{e^{i (\vec{Q} \cdot x + Q_0 t)}}{Q^2+s} \; .$$
A prominent resonance corresponds to a region of considerable
spectral strength over a relatively small area.  A caricature of
such a spectral function is shown in fig.~\ref{fig:res}

As was noted in the introduction, the prescription developed in
this paper here to isolate the effects of spontaneous chiral
symmetry breaking only works in Euclidean space.  On one hand,
this is not a serious drawback.  After all, the only practical
method to calculate hadronic quantities from QCD in an {\it ab
initio} way is via lattice QCD in Euclidean space. On the other
hand, extracting information about excited states from Euclidean
space calculations is intrinsically hard---at least
numerically---and extracting information about resonances is
harder. Fortunately, for the purpose of establishing the
theoretical result that the notion of effective chiral restoration
is well defined, it is sufficient to show that properties of
hadronic resonances can be computed from Euclidean space
correlation functions as a matter of principle. Thus, the central
issue of principle is whether the spectral functions can be
extracted given knowledge of the Euclidean space correlation
functions. Since the correlation function is an integral
transform of the spectral function, the problem amounts to
finding the inverse of the transform.  The problem is easily
converted to a standard one of the inverse Laplace transform by
focusing on the spatial integral of the correlation function at
fixed time:
\begin{equation}
\Pi(t) \equiv \int d^3 x \, \Pi(\vec{x},t) \,= \, \int d s \,
\frac{\rho(s)}{2 \sqrt{s}}\,  e^{- \sqrt{s} t} \; .
\end{equation}
Unfortunately, extracting inverse Laplace transform is a notoriously
difficult problem---indeed, it is known to be numerically unstable.
However, given sufficiently accurate computations for spectral
functions there are methods which produce spectra that can be made
to closely approximate the correct one to as much accuracy as one
wishes. These techniques suffer from an important practical
defect---namely, that the numerical difficulty of computing at fixed
level of accuracy grows exponentially as one goes up in the spectrum
and the difficulty of computing at an increased level accuracy at a
fixed $s$ also grows exponentially with accuracy. However for the
issue of principle it is sufficient to know that it is computable.

A typical method is straightforward: one characterizes the spectral
function by some functional form with a fixed number of parameters,
$N$. One calculates $\Pi (t)$ at $N$ distinct values of $t$ chosen
according to a prefixed scheme.  One then fits the parameters to
ensure the correct value of the correlation function at theses
points.  The spectral function is increasingly well described as $N
\rightarrow \infty$.

A particularly useful way to parameterize the spectral function is
via $N/2$ poles, each of variable strength.  Note that for any
finite value of $N$ this will be a very poor approximation of the
spectral function in a point-by-point sense.  However, this is not
relevant; for a sufficiently low part of the spectrum it will
accurately reproduce the integrals of spectral strength over regions
large compared with the typical level spacing.   One can replace the
spectral function constructed by the poles by a histogram. As $N$
gets large, such histograms will approximate the spectral function
with increasing accuracy.  If $N$ is large enough there will be many
poles over the characteristic scale of the resonances and one can
then resolve the resonant structure. Note that exact calculations of
correlation functions on systems of finite spatial extent {\it
always} yield discrete poles. As the volume becomes large the poles
become dense. It is relatively easy to extract the position and
residue of the lowest pole but it becomes exponentially harder to
extract poles as one goes up in energy.  Nevertheless with
sufficient numerical precision one can extract as many poles needed
to probe the spectral function in a given region.

Thus, given sufficiently accurate Euclidean correlation functions
one can reconstruct the spectral functions well enough to resolve
resonant structures with as much accuracy as we wish.

\section{Spontaneous chiral symmetry breaking and modes of zero virtuality}

Euclidean space correlation functions in QCD for currents may be
written as a functional integral over gluonic fields.  The structure
may be quite complicated and may have many terms. However, all terms
are contractions (in color, flavor and Dirac index) of structures of
the following form:
\begin{equation}
\frac{\int  D[A] \, E^{-S_{YM}} \prod_{f} Det \left( \Dslash +
m_f \right ) f[A] \prod_{k=1}^{k_{\rm max}} G_A}{\int  D[A] \,
E^{-S_{YM}} \prod_{f} Det \left( \Dslash + m_f \right )}
\label{fun}
\end{equation}
where $A$ is a gluon background configuration, $S_{YM}$ is the
Yang-Mills action for the configuration, $f$ indicates flavor,
$G_A=(\Dslash + m) $ is the quark propagator in the presence of the
gluon background field $A$, $k_{max}$ indicates the total number of
quark propagators in the term and $f(A)$ is associated with any
explicit gluon fields in the current.  We are interested in  chiral
symmetry and hence in the $m_f \rightarrow 0$ limit.

Chiral invariance is the decoupling of the left-handed and
right-handed quark fields.  The only term in the QCD Lagrangian
which couples left-handed and right-handed quarks are the mass
terms.  Chiral symmetry breaking is the coupling of left-handed and
right-handed quarks in the limit of $m_q \rightarrow 0$. The
observable chiral symmetry breaking effects in correlation functions
necessarily arise from the coupling of left-handed fields and
right-handed fields in the external currents. This in turn means
that if we can construct currents which do not couple left-handed
and right-handed quarks, we will suppress all effects of chiral
symmetry breaking.

In terms of Eq.~(\ref{fun}) the only way left-handed and
right-handed quark fields couple is through the propagators $G_A$.
One can always decompose the propagator into a chiral conserving
part (which couples left-handed quarks to left-handed quarks and
right-handed quarks to right-handed quarks) and a chiral breaking
part which couple left-handed quarks to right-handed quarks.
\begin{eqnarray}
G_A & = & G_A^{c} + G_A^{ b}  \label{gb}\\
\nonumber\\ G_A^{ c} &=& \frac{(1 + \gamma_5) G_A (1- \gamma_5)
\, + \, (1 - \gamma_5) G_A (1 + \gamma_5)}{4} \nonumber \\
\nonumber\\ G_A^{ b} &=& \frac{(1 + \gamma_5) G_A (1+ \gamma_5)
\, + \, (1 - \gamma_5) G_A (1 - \gamma_5)}{4} \nonumber \;.
\end{eqnarray}

Before proceeding it is useful to illustrate the general results
with a specific example.  Consider the vector and axial-vector
isovector currents (associated with the $\rho$ and $A_1$
channels):
\begin{equation}
J^{\mu \, a} \equiv \overline{q} \gamma^\mu \tau_a q \;\;\;J^{\mu
\, a}_A \equiv \overline{q}\gamma_5 \gamma^\mu \tau_a q
\end{equation}
These two currents together form a $(3,0)+(0,3)$ chiral-parity
multiplet; they transform into each other under axial rotations.
It is straightforward to see that
\begin{widetext}
\begin{eqnarray}
\Pi^{\mu \nu} (x) & \equiv& \langle J^{\mu \, a} (x) J^{\nu \,
a}(0) \rangle = 4 \langle \langle \gamma^\mu G^c(x,0) \gamma^\nu
G^b(0,x) \rangle \rangle + 4 \langle \langle \gamma^\mu G^b(x,0)
\gamma^\nu
G^c(0,x) \rangle \rangle  \nonumber \\
\Pi_A^{\mu \nu} (x) &  \equiv& -\langle J_A^{\mu \, a} (x)
J_A^{\nu \, a}(0) \rangle = 4 \langle \langle \gamma^\mu G^c(x,0)
\gamma^\nu G^c(0,x) \rangle \rangle - 4 \langle \langle
\gamma^\mu G^c(x,0) \gamma^\nu G^c(0,x) \rangle \rangle
\label{VA}\end{eqnarray}
\end{widetext}
where the $\langle \langle \rangle \rangle$ notation is shorthand
for averaging over gluonic configurations weighted by
$e^{-S_{YM}} \prod_{f} Det \left( \Dslash + m_f \right )$. The
critical issue illustrated here is that the {\it only} thing
separating these correlation functions is the contributions from
the chiral breaking parts of the correlators.  This property is
generic.

We ultimately wish to derive a Banks-Casher\cite{BC} type relation
in which we show  that in the appropriate limit all of the chiral
symmetry breaking effects come from states with zero virtuality.
Thus it is instructive to write these propagators in terms of
eigenmodes of the Dirac operator. To do this we first put the system
into a finite space-time box by imposing appropriate boundary
conditions.  This causes the eigenmodes to be discrete. We will take
the infinite volume limit at a later stage.  The propagator can then
be expressed as
\begin{equation}
G_A(x,x') = \sum_j \frac{\psi_j(x) \psi^\dagger_j(x')}{i \lambda_j
+ m} \label{GA} \end{equation} where $\psi_j(x)$ is an
eigenfunction of the Dirac operator $\Dslash$ with eigenvalue $i
\lambda_j$ (color, flavor and dirac indices have been suppressed).
Combining Eq.~(\ref{GA}) with Eq. (\ref{gb}) one can write the
chiral-conserving and chiral-breaking propagators in terms of
density matrices composed of the modes:
\begin{widetext}
\begin{eqnarray}
G_A^{c} & = & \int_{- \infty}^\infty  d \lambda \, \frac{
\rho^c_A(x,x':\lambda)} {i \lambda + m } \; \;
\; G_A^{ b}  =  \int_{- \infty}^\infty  d \lambda \, \frac{\rho^b_A(x,x';\lambda)}{i \lambda + m }  \\
\nonumber\\ \rho^c_A(x,x'; \lambda) &=&  \, \sum_j \, \delta
(\lambda-\lambda_j) \, \frac{(1 + \gamma_5) \psi_j(x)
\psi^\dagger_j(x') (1- \gamma_5) \, + \, (1 - \gamma_5)  \psi_j(x)
\psi^\dagger_j(x') (1 + \gamma_5)}{4}  \nonumber \\
\nonumber\\  \rho^b_A (x,x':\lambda) &=&   \,\sum_j \, \delta
(\lambda-\lambda_j) \, \frac{(1 + \gamma_5) \psi_j(x)
\psi^\dagger_j(x') (1 + \gamma_5) \, + \, (1 - \gamma_5)
\psi_j(x) \psi^\dagger_j(x') (1 - \gamma_5)}{4}  \nonumber\;.
\label{rhodef} \end{eqnarray}
\end{widetext}

It is straightforward to derive some general properties of the
propagator using the spectral decomposition. Since $\Dslash$
anti-commutes with $\gamma_5$, every eigenfucntion $\psi_j$ with a
non-zero eigenvalue $i \lambda_j$
has a partner $\gamma_5 \psi_j$
which is also an eigenvector and has eignevalue $- i \lambda_j$;
thus non-zero modes come in pairs. Zero modes are special: one can
have unpaired zero modes provided $\gamma_5 \psi = \pm \psi$.  Thus
unpaired zero modes always have fixed chirality either left or
right.  This in turn means that the unpaired zero modes do not
contribute to $\rho^c$ but can contribute to $\rho^b$. Using this
fact plus  paired structure of the eigemodes it is easy to see that
\begin{eqnarray}
\rho_A^c(x,x';\lambda)(x,x') & = & - \rho^c(-\lambda)(x,x') \;
\nonumber \\ \rho_A^b(x,x';\lambda) & = & \rho^b(x,x';-\lambda)
\; .
\end{eqnarray}
Thus the chiral conserving and chiral breaking parts of the
spectral function can be written as
\begin{eqnarray}
G_A^{ c} & = &  \int_{- \infty}^\infty d \lambda \, \,
\rho^c_A(x,x';\lambda) \,
\frac{ - i \lambda }{ \lambda^2 + m^2 } \nonumber \\
G_A^{ b} & = &  \int_{- \infty}^\infty  d \lambda \, \,
\rho^b_A(x,x';\lambda) \, \frac{ m }{ \lambda^2 + m^2 }
\label{G}\end{eqnarray}

We are interested in the chiral limit of $m \rightarrow 0$.
Naively it appears that $G_A^c$ will be nonzero in this limit
while $G_A^b$ automatically vanishes.  Of course, this is not
quite right.  In order for spontaneous symmetry breaking to occur
one must take the limit $V \rightarrow \infty$ prior to the $m
\rightarrow 0$ limit. On can proceed rather simply in a formal
manner.  First take the $V \rightarrow \infty$ limit.  This
renders the $\rho^b$ and $\rho^c$ as continuous functions of
$\lambda$  rather than as a discrete sum over delta functions in
Eqs.~(\ref{rhodef}). One can then take the $m \rightarrow 0$
limit  using the fact that $\lim_{m \rightarrow 0}
\frac{m}{\lambda^2 + m^2} = \pi \delta (\lambda)$:
\begin{eqnarray}
G_A^{ c}(x,x') & = &  \int_{0}^\infty  d \lambda \, \,
 \,
\frac{ - 2 i \rho^c_A(x,x';\lambda)}{ \lambda } \\
G_A^{ b} (x,x')& = &  \, \pi  \rho^b_A(x,x'; 0) \, \;.
\label{Gf}
\end{eqnarray}
Equation (\ref{Gf}) may be regarded as a
generalized Banks-Casher formula.  The key point here for our
purposes is that $G_A^{ b} (x,x';\lambda)$ {\it only} gets
contributions from the neighborhood of zero virtuality in the
infinite volume and chiral limits while $G_A^{ c} (x,x';\lambda)$
gets no contributions from modes approaching zero virtuality.

This generalized  Banks-Casher formula can be combined with the
fact that all chiral symmetry breaking effects in Euclidean
correlation functions ultimately are traced to the $G_A^{ b}$'s in
the functional integrals.  This allows one to deduce the general
result that {\it all} chiral symmetry breaking effects to zero
mode contributions.  Thus, for example if one were to consider
the case of the vector and axial correlation functions from
Eq.~(\ref{VA}) one sees that
\begin{equation}
\Pi^{\mu \nu}(x) - \Pi_A^{\mu \nu} (x) = 8 \pi^2 \langle \langle
\gamma^\mu \rho^b_A(x,0; 0)\gamma^\nu \rho^b_A (0,x;0)\rangle
\rangle \; .
\end{equation}
The difference of the vector and axial correlators is a direct
measure of an effect of spontaneous chiral symmetry breaking and it
is completely determined by the density matrices at zero virtuality.

\section{Excluding modes of near zero virtuality}

As noted above, all effects of chiral symmetry breaking in
Euclidean correlation functions are due to contributions from of
modes which go to zero virtuality in the infinite volume limit in
quark propagators connected to external currents. Thus, by
excluding all of these modes---and only these modes---from a
calculation of a correlation function one is removing the effects
of chiral symmetry breaking while doing no other violence to the
dynamics. This section proposes a simple explicit construction
which allows  this to be done in a simple and transparent way
while preserving all other features of QCD.

The key point is that the dependence is only for  quark
propagators connected to external currents.  Thus if one were
able to construct external currents which couple to the modes with
non-zero virtuality in the standard way while not coupling to
modes in the neighborhood of zero one would have constructed a
direct way to probe the sensitivity of the correlation function to
chiral symmetry breaking.

The basic tool in constructing such currents is a non-local quark
field operator defined as
\begin{equation}
q^\spadesuit (x; \lambda_0) \equiv \theta \left ( -\Dslash^2 -
\lambda_0^2 \right ) q(x)
\end{equation}
where $\theta$ is the standard step-function and the spade notation
indicates that the contribution from all modes whose virtuality has
a magnitude less than $\lambda_0$ have been ``dug out''.  Ultimately
the limit of $\lambda_0 \rightarrow 0$ will be taken so that only
the immediate neighborhood of zero virtuality will be excluded. One
disadvantage of this field operator is that it is non-local.
However, this does not prevent it being used to compute correlation
functions. A second disadvantage is that there is no obvious way to
interpret the field in Minkowski space. However, this should not be
viewed too negatively---the entire issue of zero modes is only
relevant in Euclidean space. More to the point the calculation of
correlation functions based on currents constructed from these
operators are well defined in Euclidean space.

The transformation properties of $q^\spadesuit (x; \lambda_0)$ under
gauge transformations is simple: they transform the same way as
ordinary quark fields;
\begin{equation}
q^\spadesuit (x; \lambda_0) \rightarrow U(x) q^\spadesuit (x;
\lambda_0) \; . \label{tran}
\end{equation}
This is easy to see.
Recall that $\Dslash \rightarrow U(x)\Dslash U^\dagger(x)$ which
implies that  any function of $\Dslash$ transforms according to
$f(\Dslash) \rightarrow U(x)f(\Dslash) U^\dagger(x)$.  Thus
\begin{equation}
\theta \left ( -\Dslash^2 - \lambda_0^2 \right ) q(x) \rightarrow
U(x) \theta \left ( -\Dslash^2 - \lambda_0^2 \right )U^\dagger(x)
U(x) q(x) \nonumber
\end{equation}
from which Eq.~(\ref{tran})
immediately follows.

Since $q^\spadesuit (x; \lambda_0)$ transforms in the same way as
$q$ under gauge transformations, for every gauge-invariant current
constructed from ordinary quark fields is a corresponding gauge
invariant current constructed from the $q^\spadesuit$. So, for
example, one can define a vector-isovector current out of these
fields:
\begin{equation}
J^{\spadesuit \mu \, a}(x;\lambda_0) \equiv
\overline{q}^\spadesuit(x;\lambda_0) \, \gamma^\mu \tau_a \,
q^\spadesuit(x;\lambda_0)
\end{equation}
Similarly one can construct correlation functions using these spaded
currents . We will denote these with a spade.

Generically these spaded-correlators can depend on the volume, the
quark mass and $\lambda_0$.  We wish to study the infinite volume,
zero mass and $\lambda_0 \rightarrow 0$ limits of these. Clearly the
ordering of the limits matters.  We will always take the infinite
volume limit first.  If we take the $\lambda_0 \rightarrow 0$ limit
first, one reproduces the correlator for the standard current in the
chiral limit
\begin{equation}
\lim_{m \rightarrow 0 \, \lambda_0 \rightarrow 0}
\Pi^\spadesuit(x; m, \lambda_0) = \lim_{m \rightarrow 0} \Pi(x,m)
\equiv \Pi(x)
\end{equation}
where $\Pi$ indicates a generic correlation function in Euclidean
space.  In contrast, if one first takes the $m \rightarrow 0$
limit and then $\lambda_0 \rightarrow 0$ limit,
\begin{equation} \lim_{ \lambda_0 \rightarrow 0 \, m \rightarrow 0 }
\Pi^\spadesuit(x; m, \lambda_0) \equiv \Pi^\spadesuit (x) \; ,
\end{equation}
one has completely removed all effects of chiral symmetry breaking
while maintaining all chiral conserving effects.

Finally we write correlators as integral transforms of spectral
functions.  For the case of two-point functions it is given by
\begin{eqnarray}
\Pi (x)  & \equiv & \langle J^\dagger (x) J(0) \rangle = \int d s
\, \rho(s) G(x;s) \nonumber \\ \Pi^\spadesuit (x)  & \equiv&
\langle J^\dagger (x) J(0) \rangle = \int d s \,
\rho^\spadesuit(s) G(x;s) \; .
\end{eqnarray}
As noted earlier this step is highly nontrivial numerically due to
the difficulty in inverting the transform, but with sufficient
precise correlation functions it can be done.  Before proceeding it
is worth noting here that the non-local nature of the fields in the
spaded currents  means that usual positivity bounds on the spectral
function need not apply.

We have reached the final stage of the construction.  In
principle, provided we have accurate Euclidean space correlators,
we now know how to compute the spectral functions with the chiral
symmetry breaking effects included or excluded.  To the extent
that the spectral functions are dominated by the chiral
conserving dynamics with $s$ in a region interest, the system is
in the regime of effective chiral restoration.  One can set the
precise criteria for identifying effective chiral restoration
based on how dominant one insights the chiral conserving parts
must be.  In regions where resonance with fixed quantum numbers
are not close together compared to typical hadronic scales a
reasonable criterion is  $\rho(s) >> \rho^\spadesuit(s)$ and
$|\partial_s \rho(s)|
>> |\partial_s \rho^\spadesuit(s)|$.

One might worry that the scheme alters the dynamics too violently
to be useful.  A particular concern is that the removal of the
zero modes might somehow undo the dynamics of confinement and
kill the resonant structures in the spaded spectral functions.
However, this cannot be the case.  This can be seen clearly by
looking at the vector and axial vector spectral functions.  As
seen in this section, the sum of the two spectral functions is
identical to the sum of the spaded spectral functions while the
differences are zero. If a resonance exist in either channel it
will appear in the sum of the two and be present in the spaded
spectral function.

Thus, as a matter of principle it is clear how to test the extent to
which effective chiral restoration occurs in QCD.  However, the real
practical question of interest is not whether a regime of
approximate effective chiral restoration occurs in the spectral
functions. On very general grounds one expect this to happen at
sufficiently large large $s$ \cite{CG,Sh}. The central question is
whether this regime occurs at low enough $s$ so that individual
resonances are discernible. Direct calculations of $\rho(s)$ and
$\rho^\spadesuit(s)$ from QCD would enable this question to be
answered.

\section{Effects of $U(1)_A$ breaking}

There is a  theoretical complication with the construction given
above. The trick was to remove all contributions in the vicinity of
zero virtuality and thereby remove all effects of chiral symmetry
breaking which can contribute in the $m \rightarrow 0$ limit, {\it
i.e.}, all effects of spontaneous chiral symmetry breaking. As
advertised in previous sections, the removal of these modes
eliminates the effects of chiral symmetry breaking and only the
effects of chiral symmetry breaking. However, these modes do more
than simply contribute to spontaneous $S(2)_L \times SU(2)_R$ chiral
symmetry breaking---they also are responsible of $U(1)_A$ breaking
effects. Recall that the $U(1)$ chiral symmetry is broken both
spontaneously and anomalously.

Ideally one would like an algorithm to  turn off the dynamical
effects of $S(2)_L \times SU(2)_R$ chiral symmetry breaking and
$U(1)$ chiral symmetry {\it separately} in order to isolate the
individual effects.  At first blush it might seem that it should
be straightforward to do this since the two chiral symmetries are
different. As will be discussed in this section, this is not
necessarily possible.

The reason that one cannot easily turn off the effects separately
is quite simple: observables associated with $U(1)$ axial symmetry
breaking may also be associated with $S(2)_L \times SU(2)_R$
chiral symmetry breaking.  Consider as a simple example the chiral
condensate: $\langle \overline{q} q \rangle$.  Clearly a nonzero
chiral condensate violates $S(2)_L \times SU(2)_R$ symmetry since
under an axial rotation in the $a$ direction $\langle
\overline{q} q \rangle \rightarrow \langle \overline{q} i
\gamma_5 \tau^a q \rangle$; but it also violates the $U(1)$
chiral symmetry since under $U(1)_A$ , $\langle \overline{q} q
\rangle \rightarrow \langle \overline{q} i \gamma_5  q \rangle$.
Thus {\it any} method which decouples the effect of the chiral
condensate in order to remove $S(2)_L \times SU(2)_R$ breaking
effects must do violence to the $U(1)$ chiral symmetry.  This is
sufficient to show that one cannot generally separately remove
the effect of $S(2)_L \times SU(2)_R$ independently from $U(1)_A$.

In fact, this connection is rather widespread.  An important
class of examples are the correlators of baryon currents.  It
turns out that when one constructs a chiral-parity multiplet of
baryon currents, states of opposite parity are connected by both
$SU(2)_A$ and $U(1)_A$ operators. This in turn implies that when
effective $S(2)_L \times SU(2)_R$ restoration occurs for these
observables, one simultaneously has $U(1)_A$ restoration. Thus as
a very practical matter it is meaningless to ask what happens if
the effects of $SU(2)_L \times SU(2)_R$ spontaneous symmetry
breaking are dcoupled while effects of $U(1)_A$ breaking are not.
Of course, consideration of the $U(1)_A$ current is complicated by
the fact that the current is broken both anomalously and
spontaneously.  From the preceding argument it is clear that for
baryon multiplets containing no isoscalars the effects of {\it
both} the spontaneous and anomolous $U(1)_A$ breaking must become
small for effective $SU(2)_L \times SU(2)_R$ restoration to
occur.  For case such as these, it is clearly pointless to try to
separate the two effects.

Of course, more generally there are cases where the two effects
are distinct.  The strategy adopted here is conservative. Namely,
the effects of both anomalous and spontaneous $U(1)_A$ breaking
and of the spontaneous breaking  $SU(2)_L \times SU(2)_R$ are
suppressed. Clearly if one sees effective restoration with this
prescription, then restoration of $SU(2)_L \times SU(2)_R$ has
occurred; the converse, however, need not be true.  Fortunately,
there is phenomenlogical evidence for the onset of effective
$U(1)_A$ restoration along with $SU(2)_L \times SU(2)_R$ in the
meson spectrum\cite{G2,G3,G4}.

\section{Practical implementation via lattice QCD}

The preceding argument makes clear that the extent to which
effective chiral restoration occurs can be quantified in QCD
spectral functions.  Thus, the issue of principal is settled.
Effective chiral restoration is a sensible notion in QCD.  What
is not clear is whether it occurs in a regime where discernible
resonances exist.  It also remains to be seen whether or not the
idea can be tested in practice via numerical simulations of
lattice QCD. Ideally a lattice version of the algorithm could be
implemented straightforwardly. One would take a large volume,
then calculate enough states to map out the spectral function and
then repeat the calculation with the spaded currents.  Both $m$
and $\lambda_0$ should be varied to make sure that the appropriate
limits are being simulated.

In practice, it will be a considerable time before it is possible to
implement this in a completely unambiguous way. The principle
problem is {\it not} the calculation of the correlators with the
spaded currents.  In practice this amounts to removing a set of
modes with low virtuality from the calculation of quark propagators.
Calculations of this sort are doable. Indeed, studies of this sort
have been done recently by Degrand \cite{Dg} precisely to study the
role of the modes responsible for chiral symmetry breaking on the
correlators.  In fact, these correlators reveal that the modes of
small virtuality contribute principally to the long-distance part of
the correlators. This is consistent with the idea of effective
chiral restoration for high mass states.

However, for a direct test of the idea of effective chiral
restoration one needs more than the correlators; one needs the
spectral functions.  The method to extract spectral information
from correlators is clear in principle.  One works with a finite
volume which discretizes the spectrum and extracts as many
discrete states as is numerically possible.  One standard way to
do this is via correlators of multiple sources with the same
quantum numbers \cite{Mich,Lu}. Since the discrete states become
increasingly dense as the box size increases, one would ideally
take a large size and thereby carefully fill out the spectral
function.  By varying the box size one can sweep through the
spectrum.

Unfortunately, the numerical noise grows exponentially as one one
goes to higher states. Thus, to go up to a fixed moderately high
mass region of interest, the cost of going to a large box is that
one needs exponentially accurate correlators.  So, in practice one
must settle for moderately small box sizes and moderately low-lying
excitations.   It is not  clear that present-day lattice simulations
are sufficient to extract resonances reliably.  The present state of
the art for extracting excited states from lattice QCD can be seen
in ref.~\cite{SA}.

The restriction to moderately small volume complicates the analysis
in two ways. The first difficulty with small volumes is that there
is not enough information to scan through the spectral functions.
One only has information about a few discrete values of energy. In
principle one can vary the volume (while keeping it moderately
small) and use the motion of the discrete levels as a function of
volume to sweep through the spectrum.  In practice this is likely to
be prohibitively expensive for some time to come.  More likely,
initial calculations to extract resonant states will be done with
only a couple of different volumes.  In general this not sufficient
to reproduce the spectral function even approximately. However, if a
state is found with a mass which is largely insensitive to the
choice of volume and whose coupling to the external currents is also
insensitive to the volume, it is probably safe to associate the
state with a relatively narrow resonance.   To the extent such an
algorithm can be used to reliably pick out resonant states, it can
be used to verify effective chiral restoration.  One can repeat the
same calculation using the spaded current.  The extent to which the
mass {\it and} the coupling to the external current remain unaltered
by this is a direct measurement of effective chiral restoration.
This suggests at least a crude test of effective chiral restoration
on the lattice might only be slightly more difficult than the task
of extracting resonant states in the first place.

Unfortunately, there is an additional complication caused by the
need to use relatively small volumes. This is a restriction on the
appropriate value of $\lambda_0$.   To work in a sensible way
near the appropriate limit,   $\lambda_0$ must be very small
compared to explicit chiral symmetry breaking effects while being
very large compared to effects associated with the finite size:
\begin{equation}
V m_\pi^2 f_\pi^2 \gg V \lambda_0^4 \gg 1  \; .
\label{ineq}
\end{equation}
For small $V$, these inequalities become impossible to maintain
unless the explicit chiral symmetry breaking terms become large
since a relatively large pion mass is needed. However, as the
explicit chiral symmetry breaking terms become large, one destroys
the very symmetry of interest. Of course, in lattice calculations
there is {\it always} as an interplay between the infinite volume
and chiral limits: as one takes quark masses to be small one must
take volumes to be large or the pionic tails of hadronic wave
functions will not fit on the lattice.   For typical lattice
applications the relevant condition is $V m_\pi^2 f_\pi^2 \gg 1$.
However, inequality (\ref{ineq}) is stronger: one needs to fit a new
scale between $V m_\pi^2 f_\pi^2$ and unity.

>From these considerations it seems apparent that practical tests of
the idea of effective chiral restoration will initially be
constrained to unrealistically large quark masses.  Potentially this
may undermine the point of the exercise---to test the chiral
properties of the states.  However, it should be noted that the idea
underlying effective chiral restoration---that the high-lying states
become insensitive to the dynamics of dynamical symmetry breaking
due to their scales---may well also apply to explicit symmetry
breaking.  Thus, if a test of the idea with unrealistically large
quark masses indicates a considerable degree of effective chiral
restoration, it is highly plausible that the real case will have a
higher level of effective restoration. Thus, while the absence of
convincing evidence of effective chiral restoration for lattice
calculations of some moderately low-lying hadrons will not rule out
the possibility, seeing the effect will provide strong evidence for
the idea.

In summary, this paper has shown that the notion of effective chiral
restoration is definable in QCD: the condition is that spectral
functions for spaded currents are similar to those of the usual
currents.  To the extent that lattice calculations can be used
reliably to extract resonances via the calculation of the idea ought
to be testable.

{\it Acknowledgments.}  The author acknowledges L. Ya. Glozman
for useful discussions. This work was supported by the U.S.
D.O.E.\ through grant DE-FGO2-93ER-40762.


\begin{thebibliography}{99}
\bibitem{G1} L. Ya. Glozman, Phys. Lett. B {\bf 475}, 329 (2000).
\bibitem{CG} T. D. Cohen and L. Ya. Glozman, Phys. Rev. D {\bf 65}, 016006 (2002);
Int. J. Mod. Phys. A {\bf 17}, 1327 (2002).
\bibitem{G2} L. Ya. Glozman, Phys. Lett. B {\bf 539}, 257 (2002);
{\it ibid}, {\bf 587}, 69 (2004).
\bibitem{G3} L. Ya. Glozman, Phys. Lett. B {\bf 541}, 115 (2002).
\bibitem{G4}  L. Ya. Glozman, Int. J. Mod. Phys. A., in press (hep-ph/0411281);
L. Ya. Glozman, A. V. Nefediev, J.E.F.T. Ribeiro, Phys. Rev. {\bf
D 72}, 094002 (2005).
\bibitem{Jido}D. Jido, T. Hatsuda, T. Kunihiro, Phys. Rev. Lett. {\bf 84}, 3252
(2000); D. Jido, M. Oka, A. Hosaka, Progr. Theor. Phys. {\bf 106},
873 (2001).
\bibitem{CG2}T. D. Cohen and L. Ya. Glozman,hep-ph/0512185.
\bibitem{Sh} M. Shifman, hep-ph/0507246.
\bibitem{JPS} R. L. Jaffe, D. Pirjol, A. Scardicchio, Phys. Rev.
Lett. {\bf 96} (2006) 121601.
\bibitem{JPS2}  R. L. Jaffe, D. Pirjol, A. Scardicchio hep-ph/0602010.
\bibitem{CN} L. Ya. Glozman and A. V. Nefediev  Phys. Rev. D{\bf 73} (2006)
074018.
\bibitem{BC} T. Banks and A. Casher,  Nucl. Phys. B {\bf 169} 103
(1980).
\bibitem{Dg} T. A. DeGrand, Phys. Rev. D {\bf 69}, 074024 (2004).
\bibitem{Mich} C. Michael, Nucl.~Phys.~B {\bf 259}, 58 (1985).
\bibitem{Lu} M. Lu\c"scher and U. Wolff, Nucl.~Phys.~B {\bf 339}, 222
(1980).
\bibitem{SA}K.J. Juge, A. Lichtl, C. Morningstar, R.G.
Edwards, D.G. Richards, S. Basak, S. Wallace, I. Sato, G.T.
Fleming,hep-lat/0601029; Tommy Burch, Christof Gattringer, Leonid
Ya. Glozman, Christian Hagen, Dieter Hierl, C. B. Lang, Andreas
Schäfer, hep-lat/0604019;

\end{thebibliography}
\end{document}